\documentclass[journal=jpccck,manuscript=article,layout=twocolumn]{achemso}

\setkeys{acs}{
abbreviations=true,
articletitle=true,
biblabel=plain,
chaptertitle=true,
doi=true,
email=true,
etalmode=truncate,
keywords=true,
maxauthors=100,
super=true
}

\setcitestyle{super,open={},close={}}

\usepackage[utf8]{inputenc} 
\usepackage{textcomp} 
\usepackage[T1]{fontenc} 
\usepackage[portuguese,english]{babel} 
\usepackage[version=4]{mhchem} 
\usepackage{siunitx} 
\usepackage{graphicx} 
\graphicspath{{Figures/}} 
\usepackage{geometry} 
\usepackage[format=plain,justification=justified,singlelinecheck=false,font={normalsize,bf},labelfont=bf,labelsep=space]{caption} 
\usepackage{float} 
\usepackage{natbib} 
\usepackage{setspace} 
\usepackage{xkeyval} 
\usepackage{array} 
\usepackage{booktabs} 
\usepackage{listings} 
\usepackage{lmodern} 
\usepackage{mathpazo} 
\usepackage[breaklinks,colorlinks=true,allcolors=blue]{hyperref} 
\hypersetup{breaklinks,colorlinks=true,linkcolor=blue,filecolor=blue,urlcolor=blue,citecolor=blue}
\hypersetup{colorlinks=true,citecolor=blue,urlcolor=blue}
\newcommand{\doi}[1]{\href{http://dx.doi.org/#1}{\nolinkurl{#1}}}
\usepackage{subfigure} 
\usepackage[compact]{titlesec} 
\usepackage{natmove} 
\usepackage{multirow} 
\usepackage{lipsum} 
\usepackage{calc} 
\usepackage{longtable} 
\usepackage{tabularx} 
\usepackage{placeins} 
\usepackage{tablefootnote}
\usepackage{adjustbox}
\usepackage{titletoc}
\usepackage{latexsym}
\usepackage{chemformula}
\usepackage{cancel}
\usepackage{amsmath}
\usepackage{amssymb}
\usepackage{amsfonts}
\usepackage{amstext}
\usepackage{xr}
\usepackage{cleveref}

\usepackage[none]{hyphenat}
\usepackage{microtype}
\makeatletter
\let\l@addto@macro\relax
\makeatother
\usepackage[fontsize=12pt]{scrextend}

\SectionsOn
\SectionNumbersOn
\AbstractOn

\title[TITLE]{The Role of the Adsorption of Alkali Cations on Ultrathin $n$-Layers of Two-dimensional Perovskites}

\author{Israel C. Ribeiro}
\affiliation[USP]{S{\~a}o Carlos Institute of Chemistry, University of S{\~a}o Paulo, P.O. Box $780$, $13560$-$970$, S{\~a}o Carlos, SP, Brazil}

\author{Pedro Ivo R. Moraes}
\affiliation[USP]{S{\~a}o Carlos Institute of Chemistry, University of S{\~a}o Paulo, P.O. Box $780$, $13560$-$970$, S{\~a}o Carlos, SP, Brazil}

\author{Albert F. B. Bittencourt}
\affiliation[USP]{S{\~a}o Carlos Institute of Chemistry, University of S{\~a}o Paulo, P.O. Box $780$, $13560$-$970$, S{\~a}o Carlos, SP, Brazil}

\author{Juarez L. F. Da{~}Silva}
\email{juarez_dasilva@iqsc.usp.br}
\affiliation[USP]{S{\~a}o Carlos Institute of Chemistry, University of S{\~a}o Paulo, P.O. Box $780$, $13560$-$970$, S{\~a}o Carlos, SP, Brazil}

\begin{document}


\begin{tocentry}
\centering
\includegraphics[width=0.90\linewidth]{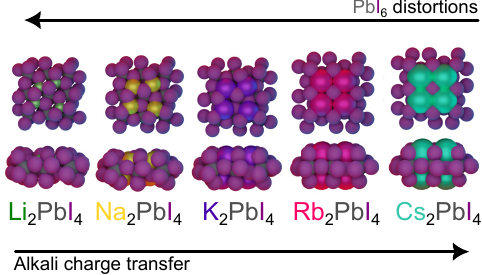}
\end{tocentry}

\begin{abstract}
Metal-halide semiconductors have great potential for real-life photovoltaic applications; however, surface defects induce several challenges to the thermodynamic stability. Here, we employed density functional theory calculations within van der Waals corrections (D3) to investigate the role of monovalent cations (\ce{Li}, \ce{Na}, \ce{K}, \ce{Rb}, and \ce{Cs}) in the passivation of 2D perovskite ultrathin films, namely, \ce{$A$2(MA_{$n$–1})Pb_{$n$}I_{$3n+1$}}, where $n = 1$ (monolayer) and $n = 2$ (bilayer). We found connections between the periodic trends of alkaline ions, such as the ionic radius, polarizability, and electronegativity, and their direct influence on the structural, energetic, and electronic properties of the slab. The increase in the distortion of the octahedra is affected by the ionic radius, and the smaller the alkaline ionic radius, the greater the out-of-phase distortion of the octahedra. Consequently, we have an increase in the band gap for systems that have greater distortions. Spin-orbit coupling effects reduce the electronic band gap while hybrid functional increases the band gap. Therefore, in most cases, the scissor operator band gap matches the value obtained from the PBE+D3 functional when comparing the final results, except for \ce{Cs} on the monolayer. The acid-base behavior of the $A$ site with the octahedra promotes a charge transfer from the alkaline to the iodide, which is more effective for systems with less electronegative alkaline ions.

\end{abstract}

\section{Introduction}

Perovskite-based materials\cite{Kojima_6050_2009} have attracted attention from the scientific community in the past \num{15} years due to their low production costs and steady increase in power conversion efficiency (PCE),\cite{Leng_908_2018,Scalon_1306_2022} indicating potential for real-life applications. The quality of perovskite thin films can be improved by controlling the physical-chemical synthesis parameters, which allows us to control the rate of nucleation and grain growth.\cite{Leng_908_2018,Ghosh_11107_2020}

The perovskite structure has the chemical formula \ce{$ABX$3},\cite{Filip_1_2014,Paritmongkol_5592_2019} where $B$ (divalent cation) and $X$ (monovalent halide-anion) atoms form the \ce{$BX$6}-octahedra, and $A$ is an intercalated monovolent cation. Thus, $A$ can be an inorganic species such as \ce{Cs} or an organic cationic molecule. The most common monovalent cations are methylammonium (\ce{MA+}),\cite{Frolova_6772_2020,Wang_16195_2016} formamidinium (\ce{FA+}),\cite{Min_749_2019,Niu_1845_2022} cesium (\ce{Cs+}),\cite{Yao_11124_2021,Huang_156_2021} or a mixture of cations.\cite{Maniyarasu_43573_2021,Frolova_131754_2021} The most common $B$ species are lead (\ce{Pb^{2+}})\cite{Rosales_906_2017,Lin_2296_2020} or tin (\ce{Sn^{2+}}),\cite{Yao_1903540_2020,Wu_863_2021} while $X$ is an anion such as iodide (\ce{I-})\cite{Park_1_2018,Filippetti_11812_2021} or bromide (\ce{Br-}).\cite{Liu_9817_2020,Man_1_2022}

The passivation of surface defects plays an important role in the quality of the light absorption layer.\cite{Li_1_2018,Jiang_1749_2019} In an attempt to reduce defects in perovskite structures, doping of monovalent alkali metal ions (\ce{Li+}, \ce{Na+}, \ce{K+}, \ce{Rb+} and \ce{Cs+}) in structures of type \ce{$ABX$3} has been extensively researched and has been shown to be quite effective.\cite{Li_284_2016,Hu_2212_2017,Hu_2018,Abdi_497_2018} For example, compared to the respective device without doping, \ce{Cs+} doping in \ce{FAPbI3} perovskite thin films ensures an increase in the thermodynamic stability of the photoactive $\alpha$-phase (cubic) and the device PCE.\cite{Li_284_2016} \ce{Rb+} doping reduces the work function ($\Phi$) in 3D/2D bilayer structures, achieving higher surface hydrophobicity that increases structural stability and PCE by over \SI{20}{\percent} in planar n-i-p devices.\cite{Tang_1253_2021} Due to their physicochemical properties, such as ionic radius and effective charge, studies show that doping with \ce{Na+} ions increases carrier lifetime and decreases trap density, showing an improvement in the performance of photovoltaic devices.\cite{Bi_1400_2017}

Thus, it has been shown that methods for passivating perovskites with alkaline cations are highly effective in improving the structural stability of these materials and increasing their tolerance to high temperatures and humid conditions.\cite{Abdi_497_2018,Ren_220_2019,Qiao_3813_2020,Xu_315504_2021,Liu_223903_2022} High levels of alkaline cations can induce the presence of second phases such as two-dimensional (2D) structures formed by the binding of alkali cations to the \ce{$BX$6}-octahedra.\cite{Kuai_1900053_2019} Passivation with high concentrations of \ce{K+} species induces the formation of 2D \ce{K2PbI4} structures, which are thermodynamically stable and effectively passivate the grain boundary and reduce the trap state. This leads to a decrease in the crystallization activation energy and contributes to increased crystallinity and grain size.

Alkali metal ions have been extensively studied in perovskite-based materials, particularly in bulk systems, through substitutional or interstitial doping.\cite{Ornelas_6607_2022} Linh et al. discovered that replacing the monovalent cation $M$ with \ce{Li+}, \ce{Na+}, and \ce{K+} in \ce{Bi_{0.5}M_{0.5}TiO3} perovskites increases the electronic band gap. This increase can be attributed to the interactions between alkali ions and oxygen ions, highlighting the crucial role of alkali ions in the anion site ($X$ site) interaction for determining band gap properties in perovskites.\cite{Linh_492_2019} Additionally, Tang et al. demonstrated that substituting the $X$ site in \ce{Cs2AgAlX6} double perovskite systems significantly affects the band gap, with a notable reduction observed in the sequence of \ce{Cl}, \ce{Br}, and \ce{I}.\cite{Tang_2046_2022} This indicates that halogen anions exert a greater influence on the band gap value compared to alkali metals.

Therefore, there is great interest in improving our understanding of the role of surface passivation of perovskites by alkaline chemical species, which can help to design better chemical species for surface passivation. In this work, we report a theoretical investigation, based on \textit{ab{~}initio} calculations on the effects of alkaline monovalent cations (\ce{Li}, \ce{Na}, \ce{K}, \ce{Rb}, and \ce{Cs}) on the structural, energetic, and electronic properties of 2D \ce{A_{2}(MA_{$n$–1})Pb_{$n$}I_{$3n+1$}} slabs, where $n = 1$ (monolayer) and $n = 2$ (bilayer). We identify an important role of alkali ionic radius in the distortion of the octahedra, which has a direct impact on the magnitude of the energetic and electronic properties of 2D ultrathin films. Furthermore, we characterize the binding mechanism and the bando ffset for the different systems.

\section{Theoretical Approach and Computational Details}

\subsection{Total Energy Calculations}

Our total energy calculations were based on the density functional theory\cite{Hohenberg_B864_1964,Kohn_A1133_1965} framework within the semilocal Perdew--Burke--Ernzerhof (PBE) formulation for the exchange-correlation energy functional\cite{Perdew_3865_1996} combined with the pairwise semi-empirical D3 van der Waals (vdW) correction proposed by Grimme.\cite{Grimme_154104_2010} The D3 correction is required to improve the description of nonlocal vdW interactions, e.g., interlayer binding energies in 2D materials,\cite{Rego_415502_2015} weak interacting adsorbates,\cite{DaSilva_085301_2007,Freire_1577_2018} hybrid-perovskites,\cite{Ozorio_3439_2020} etc. Furthermore, to overcome the limitations of PBE in describing the fundamental energy band gap ($E_g$) and the relative position of the maximum valence band (VBM) and the minimum conduction band (CBM), we employed the hybrid Heyd--Scuseria--Ernzerhof (HSE06) functional\cite{Heyd_8207_2003} for frozen DFT-PBE+D3 structures. Spin-orbit coupling (SOC) effects play a crucial role in the electronic structure of materials containing heavy elements such as \ce{Pb}, which can strongly affect the relative position of the VBM and CBM states. Therefore, SOC effects for electronic states were considered using the second-variation approach\cite{Koelling_3107_1977} for frozen DFT-PBE+D3 structures, as SOC effects produce negligible changes in the structure properties.

To solve the Kohn--Sham (KS) equations, we employed the all-electron full-potential projector augmented wave (PAW) method,\cite{Blochl_17953_1994,Kresse_1758_1999} as implemented in the Vienna \text{ab{~}initio} simulation package (VASP),\cite{Kresse_558_1993,Kresse_11169_1996} version $5.4.4$, where the KS orbitals are expanded in plane waves. For adsorbate systems, equilibrium structures were obtained by optimizing the stress tensor\cite{Francis_4395_1990} in the $xy$-plane (frozen supercell in the $z$-direction) and atomic forces for all directions without symmetry constraints using a cutoff energy that is \SI{50}{\percent} higher than the highest recommended cutoff energy considering all selected chemical species (\ce{H}, \ce{Li}, \ce{C}, \ce{N}, \ce{Na}, \ce{K}, \ce{Rb}, \ce{I}, \ce{Cs}, \ce{Pb}), i.e., $\texttt{ENCUT} = \SI{559.279}{\electronvolt}$. For the remaining properties, which are less dependent on the cutoff energy, we used a cutoff energy of $\texttt{ENCUT} = \SI{473.515}{\electronvolt}$, which is \SI{12.5}{\percent} higher than the highest recommended cutoff energy. 

For the integration of the Brillouin zones, we employed a \textbf{k}-point mesh of $2{\times}2{\times}1$ for all calculations, except for the density of states (DOS), where the \textbf{k}-mesh was increased to $4{\times}4{\times}1$ as it requires a larger number of \textbf{k}-points to obtain a well-converged DOS. Furthermore, for DOS calculations, the number of bands (\texttt{NBANDS}) was increased to be equal to the number of valence electrons. The equilibrium structures for all systems were reached once the atomic forces on each atom were smaller than \SI{2.50e-2}{\electronvolt\per\angstrom} using an energy convergence criterion of \SI{e-5}{\electronvolt}.

\subsection{Atomic Structure Models} 

\begin{figure}[t!]
    \centering
    \includegraphics[width=0.90\linewidth]{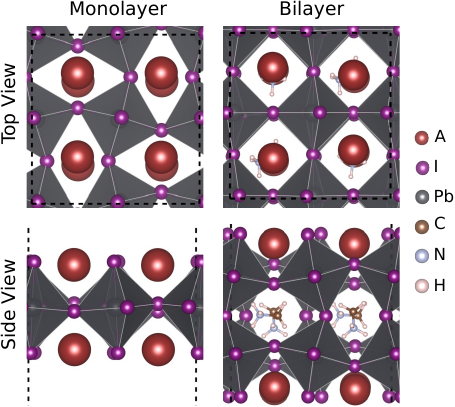}
    \caption{Top and side views of the monolayers and bilayers 2D alkali-perovskite structures, which has the \ce{$A$2PbI4} and \ce{$A$2(MA)Pb2I7} compositions, respectively (\ce{$A$} = \ce{Li}, \ce{Na}, \ce{K}, \ce{Rb}, \ce{Ce}). The dashed line represents the unit cell.}
    \label{fig:top_side_view}
\end{figure}

As stated previously, our goal is to provide an improved atomistic understanding of the surface passivation effects induced by the presence of chemical species with differing atomic radii in ultra-thin \ce{MAPbI3} films. For that, we designed slabs with initially square $2{\times}2$ surface $(100)$ unit cells composed of one (monolayer) or two layers (bilayers) of \ce{PbI6}-octahedra separated by a vacuum thickness of \SI{15}{\angstrom} to minimize interactions between the slab and its adjacent images. Therefore, based on the electron counting rule, each side of the slab (top and bottom) requires \num{4} monovalent cations ($A$) for their passivation, resulting in the following supercells: \ce{$A$8Pb4I16} (monolayer, $28$ atoms) and \ce{$A$8(MA)4Pb8I28} (bilayer, $76$ atoms), where $A$ = \ce{Li}, \ce{Na}, \ce{K}, \ce{Rb}, and \ce{Cs}. For the initial configurations, the passivating atoms were placed at the ideal positions of the $A$ cations as in the bulk \ce{MAPbI3} structure, Figure{~}\ref{fig:top_side_view}. However, due to structure optimization without symmetry constraints, the square surface $2{\times}2$ unit cell ($a_0 = b_0$) ends up as a rectangular surface unit cell after optimization ($a_0 \neq b_0$) for all calculations.

\begin{figure*}[t!]
    \centering
    \includegraphics[width=0.90\linewidth]{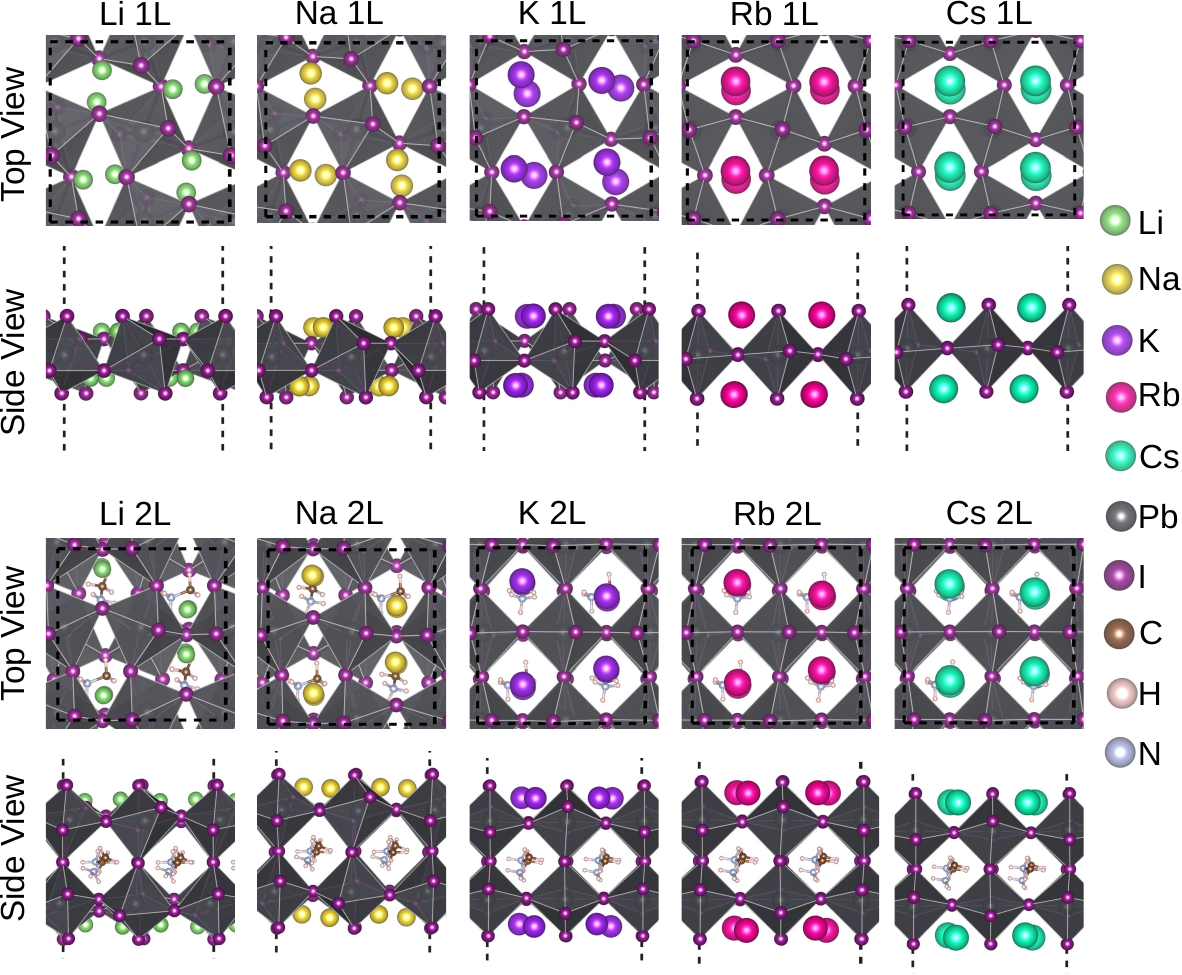}
    \caption{Top and side views of the 2D \ce{$A$2PbI4} (1L) and \ce{$A$2(MA)Pb2I7} (2L) structures in their lowest energy configurations.}
    \label{fig:structures_opt}
\end{figure*} 

\section{Results and Discussion}

To improve our atomistic understanding of the passivation effects induced by cations with different atomic radius, we calculated a wide range of physicochemical properties (descriptors) that provides a path to characterize those effects. In the following, the most important findings will be reported and discussed, while supporting data and analyses are reported within the electronic supporting information (SI) file.

\subsection{The Role of Alkali Species on the Slab Structure Deformations}

\begin{figure*}[t!]
    \centering
    \includegraphics[width=0.90\linewidth]{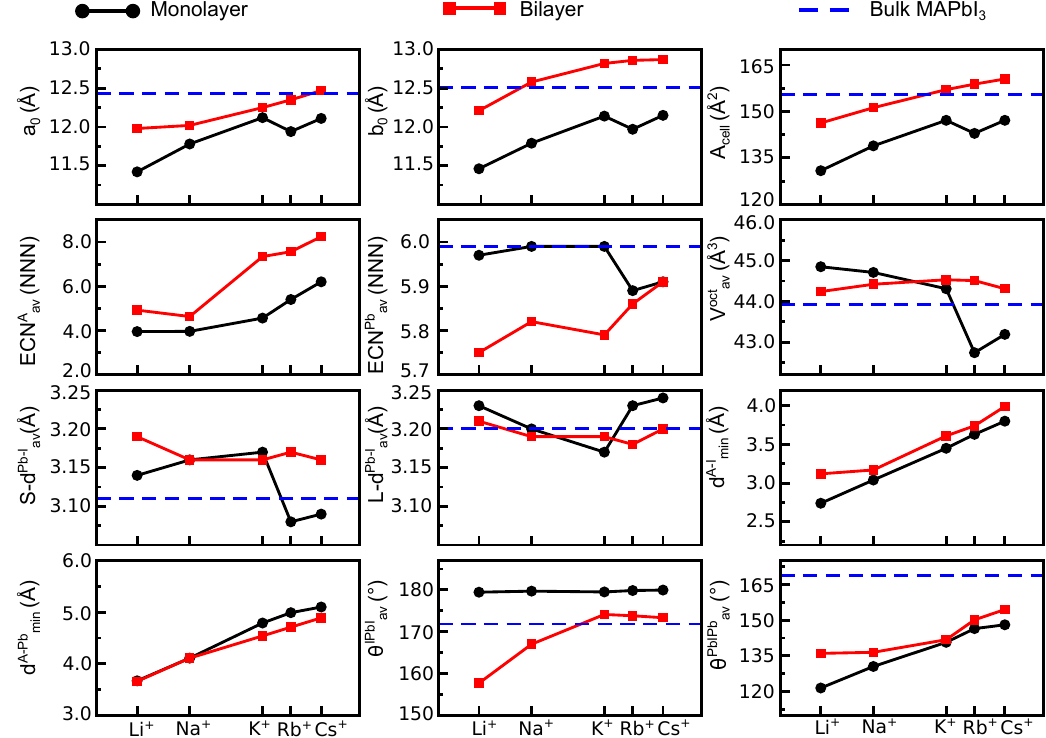}
    \caption{Equilibrium structural parameters for 2D \ce{$A$2PbI4} and \ce{$A$2(MA)Pb2I7} systems as a function of the ionic alkali atomic radius: lattice parameters ($a_{0}$, $b_{0}$), surface unit cell area ($A_{cell}$), effective coordination number for the \ce{Pb} ($ECN_{av}^{\ce{Pb}}$) and alkali ($ECN_{av}^{A}$) species, average volume of the \ce{PbI6}-octahedra ($V_{av}^{Oct}$), average bond distance of the short (S-$d_{av}^{\ce{Pb\bond{-}I}}$) and long (L-$d_{av}^{\ce{Pb\bond{-}I}}$) bonds, average minimum distance of the alkali species to the iodine and lead species ($d_{min}^{\ce{A\bond{-}I}}$, $d_{min}^{\ce{A\bond{-}Pb}}$), average angles for the \ce{I\bond{-}Pb\bond{-}I} ($\theta_{av}^{\ce{IPbI}}$) and \ce{Pb\bond{-}I\bond{-}Pb} ($\theta_{av}^{\ce{PbIPb}}$) combinations. The atomic radius are: \SI{0.90}{\angstrom} (\ce{Li^{+}}), \SI{1.16}{\angstrom} (\ce{Na^{+}}), \SI{1.52}{\angstrom} (\ce{K^{+}}), \SI{1.66}{\angstrom} (\ce{Rb^{+}}), and \SI{1.81}{\angstrom} (\ce{Cs^{+}}).}
    \label{fig:structural_parameters}
\end{figure*}

Several structural optimizations were performed, and the lowest energy configurations for each system are shown in Figure{~}\ref{fig:structures_opt}. To characterize the structural effects induced by the presence of alkali species in ultrathin \ce{MAPbI3} films, we selected a set of geometric parameters, namely: $(i)$ lattice parameters ($a_0$, $b_0$) and surface area ($A_{cell}$), $(ii)$ average effective coordination numbers for selected species (ECN$_{av}^i$), where $i$ indicates alkali species or \ce{Pb}, $(iii)$ mean volume of \ce{PbI6}-octahedra ($V_{av}^{oct}$), $(iv)$ analysis of bond lengths and $(v)$ angles. All results are shown in Figure{~}\ref{fig:structural_parameters}, and more details of the structural results are found in SI.

\paragraph{Lattice Deformations:} As expected, the lattice parameters ($a_0$, $b_0$) are strongly affected by the thickness of the slab (number of \ce{PbI6}-layers), adsorption of the alkali species, and internal distortion of the \ce{PbI6}-octahedra, which is supported by the following findings: $(i)$ There is a significant difference in the lattice constant values for one and two \ce{PbI6}-monolayers (e.g., $a_0 < b_0$ by up to \SI{5.45}{\percent}), which is almost independent of the alkali species. This effect can be explained by quantum-size effects,\cite{DaSilva_195416_2005} which yields a strong confinement in single monolayers. $(ii)$ We found an almost monotonic increase in the lattice parameters from \ce{Li} to \ce{Cs}, which can be correlated with the increasing the ionic radius of the alkali species, that is, from \ce{Li^{+}} (\SI{0.90}{\angstrom}) up to \ce{Cs^{+}} (\SI{1.81}{\angstrom}).\cite{Lopes_363_2006} $(iii)$ Differences in the values of $a_0$ and $b_0$ can be explained by the asymmetric distortions of the \ce{PbI6}-octahedra, which result in the existence of short and long \ce{Pb-I} bonds, Figure{~}\ref{fig:structural_parameters}. $(iv)$ As expected, the lattice parameters increase towards the values obtained for the bulk \ce{MAPbI3} ($a_0 = \SI{12.43}{\angstrom}$, $b_0 = \SI{12.51}{\angstrom}$) as the alkali atomic number increases, which can be explained by the ionic radius of the MA molecule, i.e., \SI{1.80}{\angstrom},\cite{Park_65_2015} which is closer to the \ce{Cs^{+}} specie (\SI{1.81}{\angstrom}).

\paragraph{\ce{PbI6}-octahedra Deformation:} Large lateral displacements (${\Delta}^{A}_{av}$) are found due to Coulombic attractions between the alkaline and iodide ions, that is, displacements in the $xy$ directions of the alkaline ions toward the iodides; therefore, alkaline ions of smaller ionic radii tend to migrate from centers of greater symmetry towards the halides, as seen in SI Figure{~}S6. This trend follows a radius/charge correlation of the alkaline ion, which increases from \ce{Li} to \ce{Cs}. The atomic \ce{$A$-I} distance increases almost linearly by increasing the cationic alkali radius, Figure{~}\ref{fig:structural_parameters}, which also affects the effective coordination number of the alkali species ($ECN_{av}^{A}$).

We found a minimum distance between \ce{Li^{+}} and \ce{I^{-}} species of \SI{2.74}{\angstrom}, which is very close to the value obtained for the lithium iodide compound (\SI{2.70}{\angstrom}),\cite{Huang_4931_2016} while the \ce{Cs-I} distance is about \SI{3.80}{\angstrom} (monolayer), i.e., close to the value obtained for the bulk \ce{CsI} (\SI{3.95}{\angstrom}).\cite{Vats_4626_2020} Thus, the atomic radius plays a critical role in the location of the alkali species and not the chemical binding mechanism.

The close proximity of the alkali ions to the halides results in out-of-phase distortions,\cite{Dias_19142_2021} where the angle between the \ce{PbI6}-octahedra ($\theta_{av}^{\ce{PbIPb}}$) is less than \SI{180}{\degree}, which characterizes an ideal angle between \ce{Pb\bond{-}I\bond{-}Pb} in a system of octahedra without distortions. Depending on the ideal angle, the monolayer \ce{Li} exhibits the highest percentage difference of about \SI{48}{\percent} and the \ce{Cs}, on the other hand, shows a deviation of \SI{16}{\percent}. Once the thickness of the layer of the \ce{PbI6}-octahedra increases, the out-of-phase distortions tend to diminish, as seen for bilayer systems, which behave closer to the bulk \ce{MAPbI3}, where we get an angle closer to the value of the ideal angle. 

The average volume of the octahedra remains mostly unchanged and the angle created by the \ce{I\bond{-}Pb\bond{-}I} bond inside the octahedra ($\theta_{av}^{\ce{IPbI}}$) remains close to the ideal angle due to the long and short bonds, i.e., there are no stretching or deformations along the \ce{Pb\bond{-}I} bonds. The increase in distortions is proportional to the radius of alkaline ions; this tendency is also found in bulk systems in general \ce{APbI3}, with $A$ representing alkaline, as reported by Park \textit{et al.}\cite{Park_1078_2019}

We found $ECN_{av}^{\ce{Pb}}$ values for monolayer systems close to the value found in the bulk \ce{MAPbI3} ($ECN_{av}^{\ce{Pb}} = \SI{5.99}{NNN}$). However, when a second layer of \ce{PbI6}-octahedra and MA interlayer molecules is added, the value of $ECN_{av}^{\ce{Pb}}$ begins to decrease, while it is still close to the value discovered for the bulk. Additionally, for bilayer systems, there is a stabilization of out-of-phase distortions towards the $a_{0}$ axis. For example, using an adaptation for the Glazer modes for 2D perovskites materials,\cite{Ming_2985_2020} we found that for the \ce{K} monolayer system, a Glazer mode corresponds to $a^{-}a^{-}$ and for its respective bilayer, we found the $a^{+}a^{-}$ mode, indicating that there are in-phase rotations for the bilayer system and out-of-phase rotations for the monolayer system.

\subsection{Adsorption and Interaction Energies} 

\begin{figure}[t!]
   \centering
   \includegraphics[width=0.90\linewidth]{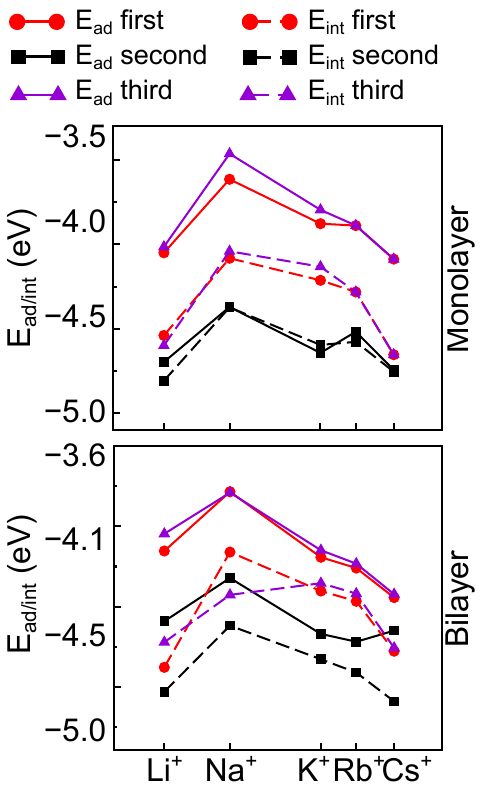}
   \caption{Adsorption ($E_{ad}$) and interaction ($E_{int}$) energies for alkaline species on the perovskite monolayer and bilayer structures. First scheme indicates the adsorption and desorption of \num{8} $A$ atoms, while the second scheme considers the adsorption of \num{8} $A$ atoms and desorption of \num{2} $A$ atoms. The third scheme is identical to the first one, however, the alkaline species were relaxed only in the $z$-direction (constraint relaxation).}
   \label{fig:Ead_Eint}
\end{figure}

To evaluate the energetic stability of the $A$ adsorbates on the perovskites thin-films, we calculated the magnitude of the interactions between the adsorbate species and perovskites slabs via the analyses of the adsorption ($E_{ad}$) and interaction energies ($E_{int}$), which can be obtained using the following equations:
\begin{align}\label{eq:Ead_first}
    E_{ad} = (E_{tot}^{2D~perovskite} - E_{tot}^{slab~relaxed} & \nonumber \\ - mE_{tot}^{A~free-cation})/m{~},
\end{align}
where $E_{tot}^{slab}$ is the total energy of the perovskite slabs, while $E_{tot}^{slab{~}relaxed}$ is the total energy of the optimized perovskite slabs through the stress tensor without the $A$ adsorbates. $E_{tot}^{A~free-cation}$ is the total energy of the free $A$-cation using a box and $m$ is the number of $A$ atoms removed from the slab.

For $E_{int}$, the following equation was used,
\begin{align}\label{eq:Eint_first}
    E_{int} = (E_{tot}^{slab} - E_{tot}^{slab{~}frozen} & \nonumber \\ + E_{tot}^{A-layer{~}frozen})/m{~},
\end{align}
where $E_{tot}^{slab{~}frozen}$ and $E_{tot}^{A-layer{~}frozen}$ are the total energies of the frozen slab and $A$-layer, respectively. To obtain additional insights, both $E_{ad}$ and $E_{int}$ were calculated using three different schemes, namely, $(i)$ adsorption and desorption of \num{8} $A$ atoms on the slab (\num{4} on each side of the slab), $(ii)$ adsorption of \num{8} atoms and desorption of \num{2} atoms, and $(iii)$ adsorption and desorption of \num{8} $A$ atoms on the slab relaxed under structural constraints, i.e., the $A$ atoms were relaxed only in $z$-direction, while all remaining atoms were relaxed without constraint. 

All results are reported within Figure{~}\ref{fig:Ead_Eint}, which yields the following conclusions: $(a)$ Except for small differences in the interaction energy, the first and third schemes produce very similar results for all systems, i.e., the lateral relaxations of alkaline species are not critical for $E_{ad}$ and $E_{int}$. Furthermore, the equilibrium lattice constants are nearly the same for both schemes, which is consistent with the trends. $(b)$ The adsorption and interaction energies are nearly identical using the scheme $(ii)$ for monolayers, while there are large differences for bilayers, indicating a larger role for slab relaxations. $(c)$ The trends in the adsorption/interaction energies are nearly the same as a function of the $A$ species, i.e., they do not depend on the calculation framework or number of layers. Basically, the adsorption/interaction energy decreases in magnitude from \ce{Li} to \ce{Na}, which then increases monotonically from \ce{Na} to \ce{Cs}, which is explained in the following. It is crucial to emphasize that the large values of $E_{ad}$ and $E_{int}$ can be attributed to the strong electrostatic interaction between negatively charged iodine atoms and positively charged alkali cations, resulting from the ionic bonding in the perovskite structure.\cite{Zhang_34402_2020}

According to the principle of hard and soft acids and bases (HSAB),\cite{Pearson_3533_1963} all compounds exhibit a tendency toward stability, with the exception of the \ce{Li} system, due to the higher \ce{PbI6}-octahedra distortions observed for this system. The HSAB principle establishes the division into small radius ions and high positive charge, defined as hard, and ions that have small radius ions and low charges, defined as soft, tends to be supported by the data $E_{ad}$ and $E_{int}$ for the \ce{Na}, \ce{K}, \ce{Rb} and \ce{Cs} systems. Hard acids prefer to bind to hard bases, and the opposite is true for soft acids. In this way, we have a stabilization of energy when we decrease the hardness of the ions, from \ce{Na} to \ce{Cs}, since the iodide ion has characteristics of a soft ion.

\subsection{Binding Mechanism Among the Alkali-slab}

To improve our atomistic understanding of the interaction mechanism of alkaline species on perovskites thin films, we calculated the magnitude of the charge transfer via density derived electrostatic and chemical (DDEC) method,\cite{Manz_47771_2016,Limas_45727_2016} electron density differences, and the electron localization function, which can be used to characterize the covalent contribution to the binding mechanism.

\subsubsection{Charge Transfer} 

\begin{figure*}[t!]
    \centering
    \includegraphics[width=0.90\linewidth]{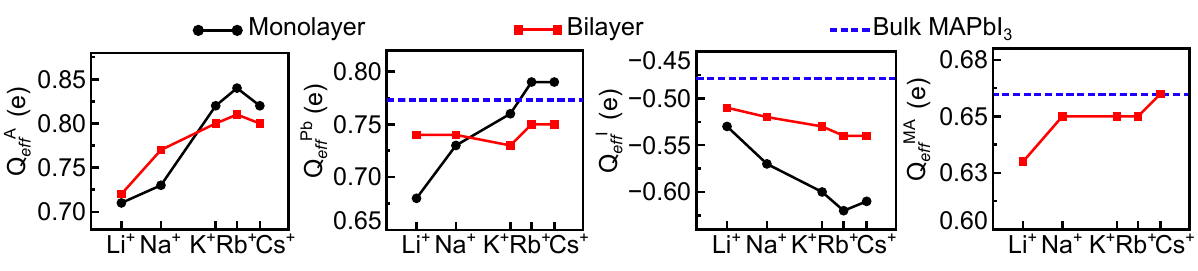}
    \caption{Effective DDEC charges for the alkaline species, lead, iodine, and MA interlayer molecules, which are indicated by $Q^{A}_{eff}$, $Q^{\ce{Pb}}_{eff}$, $Q^{\ce{I}}_{eff}$, and $Q^{\ce{MA}}_{eff}$, respectively.}
    \label{fig:charge2D}
\end{figure*}

The effective DDEC charge results are shown in Figure{~}\ref{fig:charge2D}. We found that alkaline ions tend to be more positive from \ce{Li+} to \ce{Cs+} ions when they are adsorbed on the slab. This conclusion is consistent with the electronegative nature of alkaline ions, as more electronegative ions have lower effective charges, indicating less electronic donation to the slab. Charge transfer is demonstrated by the inverse proportional relationship between the charges of the alkaline ions ($Q^{A}_{eff}$) and the charges of the respective iodides  ($Q^{\ce{I}}_{eff}$), i.e., as $Q^{A}_{eff}$ increases, the values of $Q^{\ce{I}}_{eff}$ decrease, indicating charge accumulation in \ce{I-}.

Similarly to the $Q^{A}_{eff}$ results, the effective charges on the lead cations ($Q^{\ce{Pb}}_{eff}$) in monolayer systems tend to be more positive from \ce{Li} to \ce{Cs}, indicating that both alkali and lead lose electron density to halides. Charge transfer is less obvious in bilayer systems, where organic molecules MA ($Q^{\ce{MA}}_{eff}$) act as charge donors, but $Q^{\ce{MA}}_{eff}$ and $Q^{\ce{Pb}}_{eff}$ are constant in all bilayer systems. Thus, $Q^{\ce{I}}_{eff}$ exhibits an almost constant behavior. In the bulk system, $Q^{\ce{Pb}}_{eff}$ is similar to that found in the monolayer \ce{K}, \ce{Rb}, and \ce{Cs} and the bilayer \ce{Rb} and \ce{Cs} systems, while $Q^{\ce{MA}}_{eff}$ is similar to that found in the \ce{Cs} system. However, $Q^{\ce{I}}_{eff}$ is higher than that found in the slab, approaching $Q^{\ce{I}}_{eff}$ of the \ce{Li} bilayer.

\subsubsection{Electron Density Difference} 

\begin{figure*}[t!]
    \centering
    \includegraphics[width=0.90\linewidth]{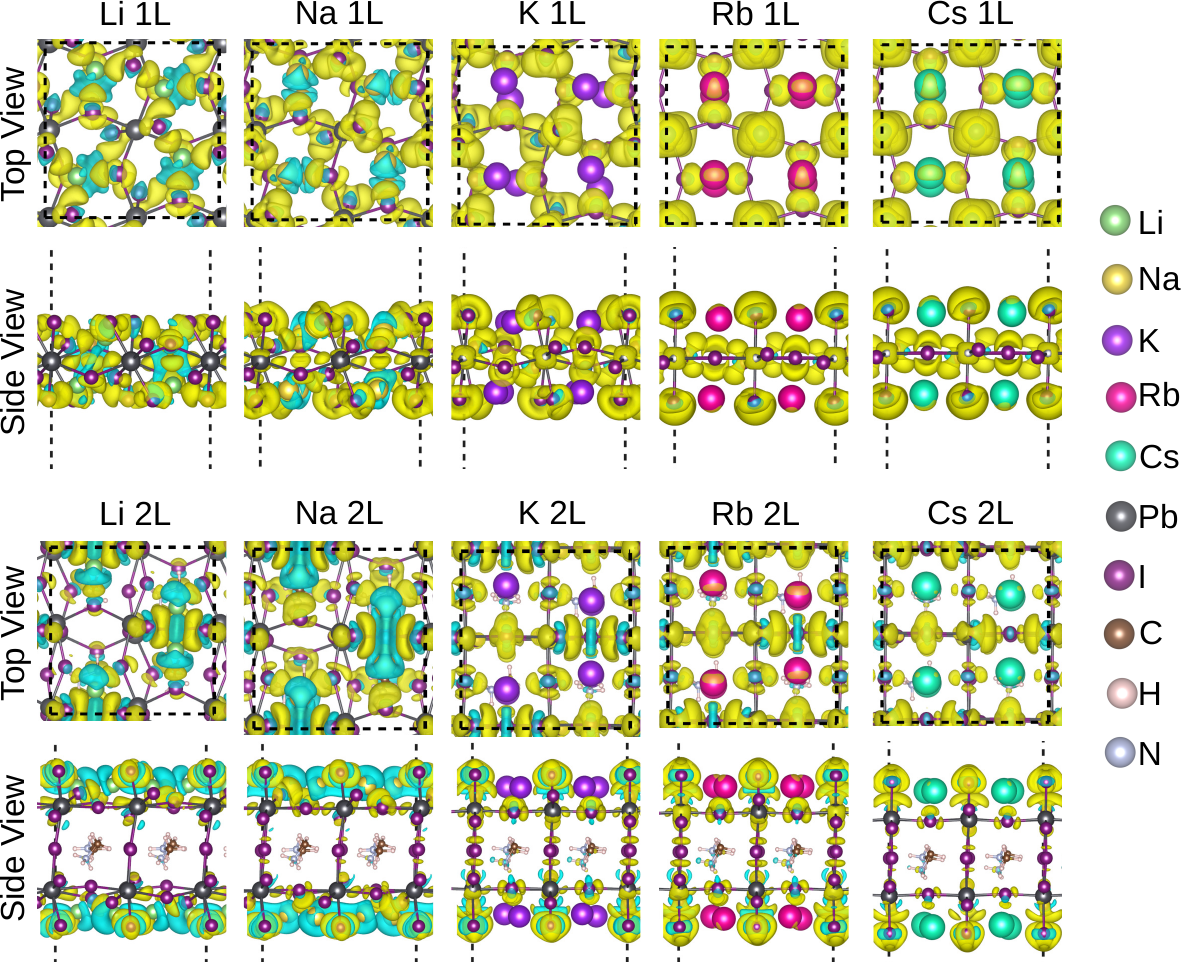}
    \caption{Electron density difference isosurfaces (\num{0.0015} $\si{bohr^{-3}}$) for the 2D perovskites-based materials. Yellow and blue regions indicate accumulation and depletion of charge, respectively.}
    \label{fig:Alkali_EDD}
\end{figure*}

To better visualize the charge transfer scenario, we performed an electron density difference analysis of the systems using the following relationship,
\begin{equation}  
\Delta\rho = \rho^{A/slab} - \rho^{A} - \rho^{slab}~,
\end{equation} 
where $\rho^{A/slab}$ is the electron density of materials based on 2D perovskites, $\rho^{A}$ is the electron density of the isolated alkaline monolayer and $\rho^{slab}$ is the electron density of the frozen 2D slab, i.e., both are fragments obtained from the \ce{$A$}/slab systems. The results are shown in Figure{~}\ref{fig:Alkali_EDD}. 

We found a large accumulation of electron charges in the neighboring \ce{I-} ions, which is expected from the DDEC analysis. Furthermore, as expected, a large charge depletion is observed in the neighboring cations of \ce{Li+} and \ce{Na+}. As a result of the charge/radius ratio of the ions, charge depletion is more evident for systems with smaller radii; however, all alkaline ions donate electrons to the iodide, as expected. Following, the effective DDEC charges, there is an increase in charge donation from \ce{Li} to \ce{Cs}, and hence the alkaline ions have a Lewis bases character, whereas the halides act as Lewis acids. Charge depletion and accumulation are also related to structural factors of compounds, such as \ce{PbI6}-octahedra distortions, with high-depletion visualization in systems with greater distortions. This result indicates that orbital overlap might be more efficient in systems with higher distortions, indicating that the adsorption of the \ce{Li+} and \ce{Na+} ions in the slab is strong.

\subsubsection{Electron Localization Function} 

We observed chemical bonds with ionic features similar to those found in highly ionic compounds of alkali metal iodides, which play an important role in the adsorption/interaction energies. However, the ionic character decreases as the atomic number of alkaline ions decreases. The \ce{Li} system exhibits the largest covalent character among all systems, as shown by the analysis of the electron localization function (ELF), because this cation has the greatest capacity to polarize the iodide electron cloud due to its charge-to-radius ratio. In fact, according to Fajan's rule,\cite{Fajans_165_1923} the covalent character in the systems is inversely proportional to the size of the cation and directly proportional to the size of the anion. As a result, the \ce{Cs+} ion exhibits a lower ability compared to alkaline ions to polarize the iodide and has a more ionic character.

\subsection{Local Density of States Analyses}

\begin{figure}[t!]
    \centering
    \includegraphics[width=0.90\linewidth]{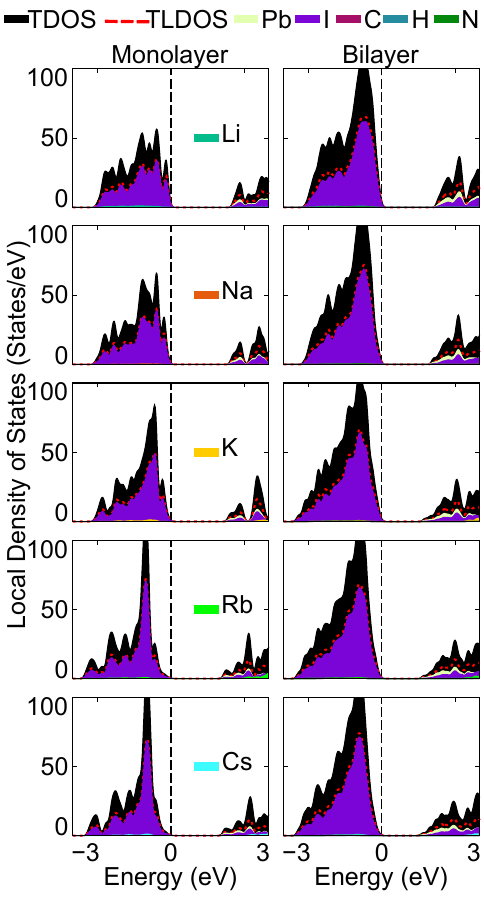}
    \caption{Local density of states (LDOS) for 2D perovskites-based materials, the valence band maximum (VBM) was set at \SI{0}{\electronvolt} (vertical dashed line).}
    \label{fig:Alkali_DOS}
\end{figure}

To move beyond, we conducted local DOS calculations to characterize the electronic states near the energy band gap window, which is critical for optoelectronic applications. The results are shown in Figure{~}\ref{fig:Alkali_DOS}. The electronic contributions near the band gap region for both monolayers and bilayers are directly derived from \ce{PbI6} atoms, which is similar to the results obtained for bulk \ce{MAPbI3}. The VBM and CBM characters are predominantly derived from the \ce{I} $s$- and $p$-states and \ce{Pb} $s$- and $p$-states, respectively. Therefore, the distortions in the \ce{PbI6}-octahedra will affect the electronic band gap, which also depends on the passivation of the alkaline species. The contribution of alkaline species to the electron density of states is minor due to the charge transfer, and hence, they are not expected to directly affect the electronic properties, but rather to improve the energetic stability.

\subsection{Spin-orbit Coupling Effects in the Band Structures}

\begin{figure*}[t!]
    \centering
    \includegraphics[width=0.90\linewidth]{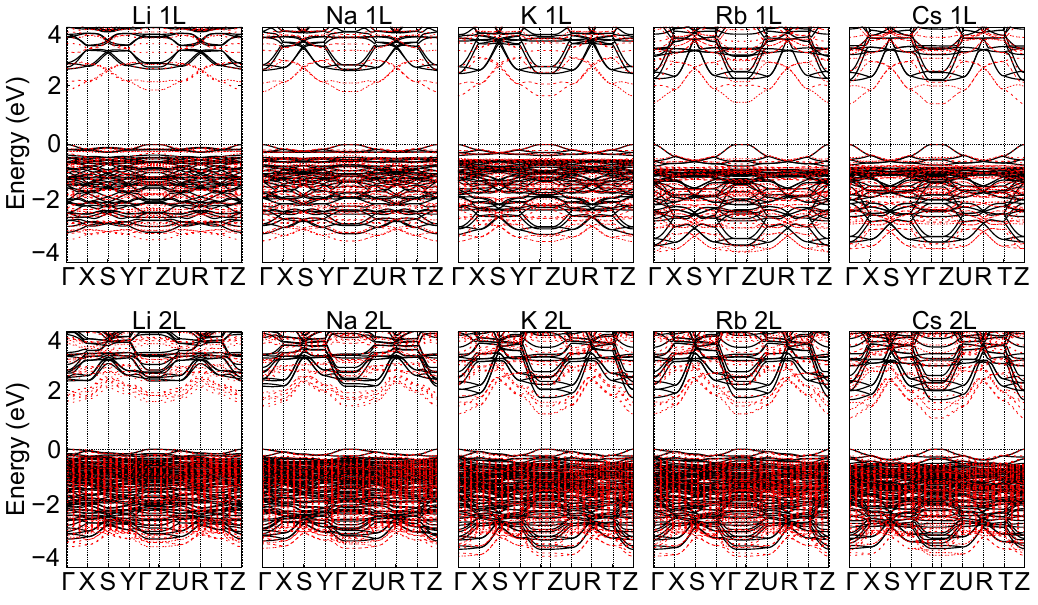}
    \caption{The electronic band structure of 2D perovskites-based materials obtained with PBE+D3 (black lines) and PBE+D3+SOC (dashed red lines) for the same equilibrium volume ($V_0$) obtained by PBE+D3 calculations. For all plots, the top of the valence band is set at \SI{0}{\electronvolt}.}
    \label{fig:bands_2DAlkPVK}
\end{figure*}

To address issues related to self-interaction errors and the underestimation of electronic band gaps when employing the PBE+D3 method, as evidenced by previous studies,\cite{Ali_49636_2020, Wang_1_2022} we have adopted an alternative approach. In this study, we aim to enhance the accuracy of fundamental band gap descriptions by implementing the HSE06 hybrid functional.\cite{Heyd_8207_2003} This choice is motivated by its anticipated ability to yield more reliable results compared to the PBE functional. To mitigate computational expenses, we have employed the scissor operator framework,\cite{Dias_085406_2020, Torres_1932_2023} which involves the incorporation of a correction derived from the spin-orbit coupling (SOC) effects and the hybrid HSE06 functional into the PBE functional. This modification allows us to estimate the fundamental band gap for these systems at the $\Gamma$-point ($E^{\Gamma}_{g}$). The outcomes of our investigation are presented in Figure~\ref{fig:bandgap}.

\subsection{Electronic Energy Band Gaps}

\begin{figure}[t!]
    \centering
    \includegraphics[width=0.90\linewidth]{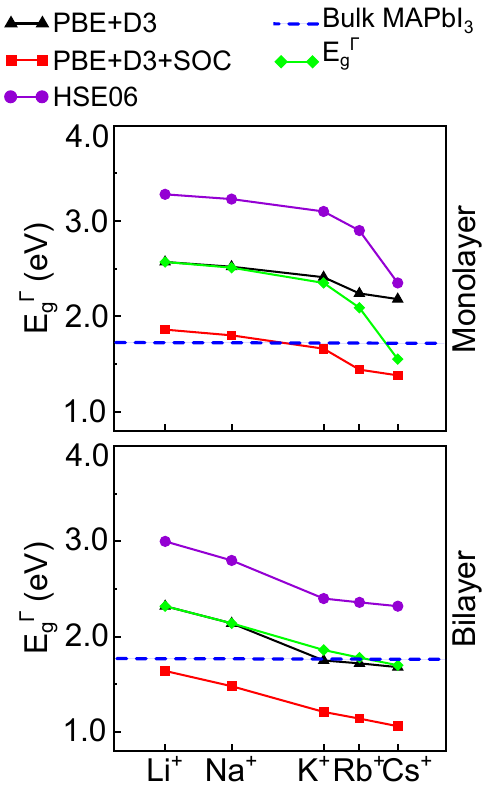}
    \caption{The electronic band gap obtained by different approximations, PBE+D3, PBE+D3+SOC, HSE06 and the fundamental gap obtained by the $E^{\Gamma}_{g}$ scissors operation, as a function of the ionic radii of the alkaline species. In addition, we report as reference the \ce{MAPbI3} bulk band gap (dashed blue line) obtained via PBE+D3 at $\Gamma$-point.}
    \label{fig:bandgap}
\end{figure}

Due to self-interaction errors, the electronic band gap computed using PBE+D3 is commonly underestimated compared to experimental results.\cite{Ali_49636_2020,Wang_1_2022} Here, to improve the description of the fundamental band gap, we use the HSE06 hybrid functional,\cite{Heyd_8207_2003} which is expected to produce better results than the PBE functional. To reduce computational cost, we used the scissor operator framework,\cite{Dias_085406_2020,Torres_1932_2023} in which a correction derived from the SOC effects and the hybrid HSE06 functional is added to the PBE functional to obtain an estimated fundamental band gap for these systems at the $\Gamma$-point ($E^{\Gamma}_{g}$). The results are shown in Figure{~}\ref{fig:bandgap}. 

We obtained the following trends: $(a)$ SOC effects reduce the electronic band gap for all systems, and the trends are the same for both PBE+D3 and PBE+D3+SOC results, i.e., from \SI{2.57}{\electronvolt} up to \SI{1.86}{\electronvolt}; $(b)$ as expected, the hybrid HSE06 functional increases the fundamental band by about \SI{0.62}{\electronvolt}, however, \ce{Cs} in single monolayers is an exception to this behavior. Thus, as expected, one effect reduces the band gap, while the second increases it. In the final result, the band gap of the scissor operator is nearly the same as that obtained by the functional PBE + D3 for most systems, except \ce{Cs} in the monolayer.

Our study found a difference in the band gap values between the functional calculations of PBE+D3 and hybrid HSE06 for the monolayer \ce{Cs} compound was only \SI{0.14}{\electronvolt}. Furthermore, the electronic band gap of the monolayer \ce{Cs} (\SI{1.38}{\electronvolt}) is comparable to the DFT band gap, with PBE+D3 considering SOC effects, for the monolayer 2D \ce{MA2PbI4} (\SI{1.40}{\electronvolt}).\cite{Zhu_3349_2018}

The systems ranging from bilayer \ce{K} to \ce{Cs} exhibit a fundamental band gap that is similar to the value calculated for the bulk \ce{MAPbI3} for similar reasons. This similarity can be attributed to the lower octahedral distortions observed in these systems, which resemble those found in the bulk \ce{MAPbI3}. These findings align with the stronger quantum confinement effect typically observed in monolayer systems, which generally leads to larger band gaps compared to their bilayer counterparts.\cite{Cheng_1_2018}

\subsection{Work Function and Band Offset Characterization}

\begin{figure}[t!]
    \centering
    \includegraphics[width=0.90\linewidth]{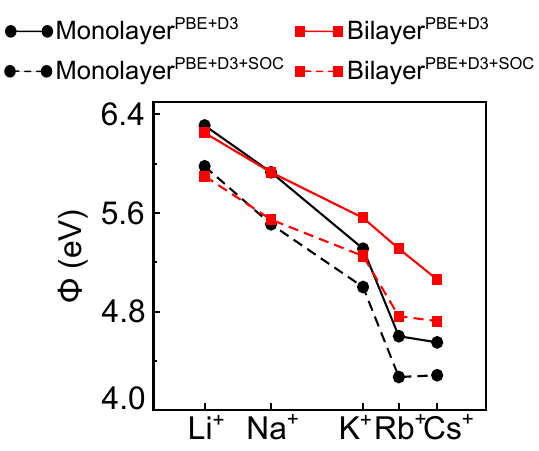}
    \caption{Work function calculated by the PBE+D3 and PBE+D3+SOC approximations for all studied systems.}
    \label{fig:workfunction}
\end{figure}

\subsubsection{Work function} 

The work function ($\Phi$) is the minimum energy required to remove an electron from the surface of a solid to the vacuum region. It can be estimated by the energy difference between the energy of the highest occupied state (VBM) and the value of the electrostatic potential in the vacuum region.\cite{Jacobs_2016,Lin_2203703_2022} The results are shown in Figure{~}\ref{fig:workfunction}. The behavior of the $\Phi$ value in the 2D alkali slab also follows a periodic trend depending on the alkali species in the systems, with an inversely proportional value to the electronegativity value of the alkali species. In other words, the \ce{Li} and \ce{Cs} systems have the highest and lowest values of $\Phi$, respectively. 

The monolayer and bilayer systems \ce{Li} and \ce{Na} have similar $\Phi$ values, but the difference increases from the \ce{K} system. This difference can be explained by the difference in electronegativity of the alkali species \ce{K}, \ce{Rb}, and \ce{Cs}.  These alkali species have values close to each other but far from the values attributed to the metals \ce{Li} and \ce{Na}. The difference in electronegativity is about \SI{21}{\percent} and \SI{16}{\percent} compared to the electronegativity of \ce{Cs}. Due to the relativistic properties of lead atoms in the systems,\cite{Pitzer_3068_1982} the $\Phi$ analyzes were carried out while considering the effects of spin orbit coupling. SOC effects directly increase the VBM energies for all systems but have no effect on the electrostatic vacuum level ($V(\textbf{r}_{vac})$). Therefore, the values of $\Phi$ decrease when the SOC effects are considered.

\begin{figure*}[t!]
    \centering
    \includegraphics[width=0.90\linewidth]{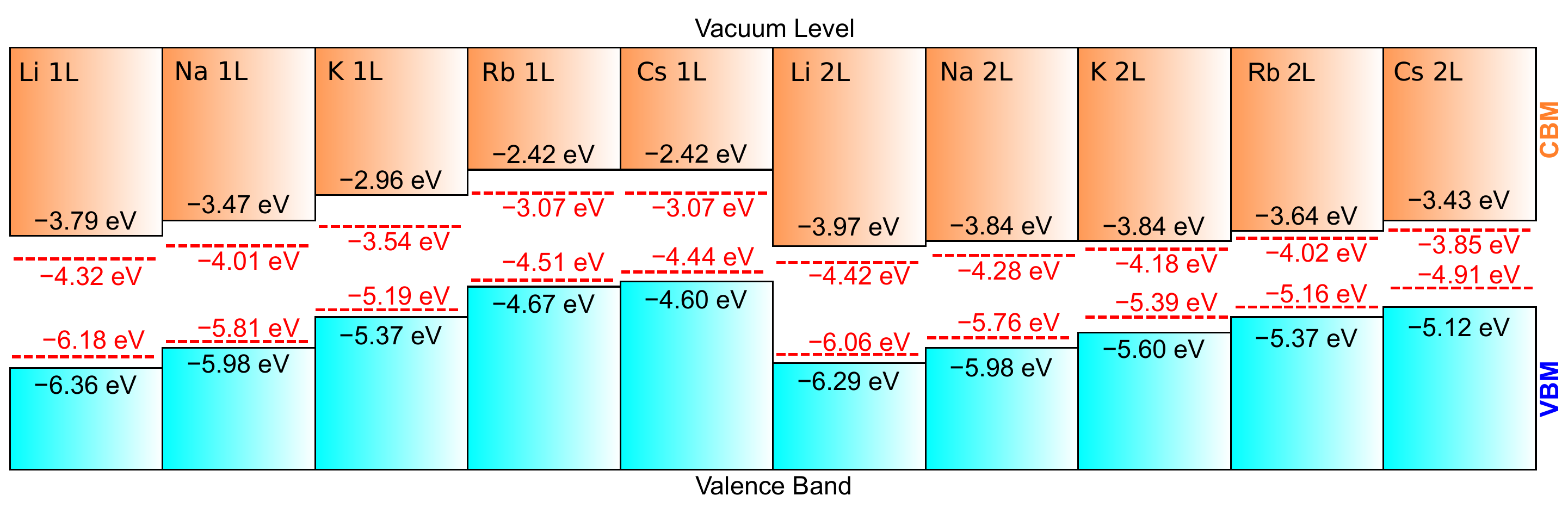}
    \caption{PBE+D3 in blue VBM and orange CBM of 2D perovskites-based materials in equilibrium lattice parameters with respect to vacuum level (zero energy), the red dashed line and values represent the effect of PBE+D3+SOC calculations on VBM and CBM.}
    \label{fig:Alkali_BOS}
\end{figure*}

\subsubsection{Band Offset Analyses} 

The electronic structures of the 2D alkali slab were aligned by adjusting their eigenvalues with respect to the vacuum level. The band offset of is shown in Figure{~}\ref{fig:Alkali_BOS} with and without SOC effects. Thus, it shows the relativistic effect affects on the CBM levels of all systems, shifting the bands to lower energy levels with a tendency to decrease levels from \ce{Li} to \ce{Cs}. In terms of VBM levels, there was a minor increase in all systems compared to changes in CBM levels. This small effect on VBM levels is even smaller in bilayer systems, where we can observe small variations between level shifts caused by SOC effects. The changes in monolayer systems range from \SI{0.53}{\electronvolt} to \SI{0.64}{\electronvolt} for the \ce{Li} and \ce{Cs} systems, respectively, while for the respective bilayer systems, the variations range from \num{0.48} to \SI{0.44}{\electronvolt}.

This observation can be explained by considering that the quantum confinement effect is smaller in bilayer systems, resulting in charge transfer stabilization and a small variation when considering SOC effects. The structural, energetic, and electronic characterizations of these materials have several similarities to those of bulk alkali metal iodide compounds. However, because of the system dimensions and, of course, the perovskite base structure, the 2D materials present exclusive properties that guarantee potential experimental studies and possible applications as 3D perovskite passivators.

\section{Conclusions}

We performed DFT calculations to investigate the role of alkaline monocations (\ce{Li}, \ce{Na}, \ce{K}, \ce{Rb} and \ce{Cs}) in the passivation of 2D \ce{$A$2(MA_{$n$–1})Pb_{$n$}I_{$3n+1$}} ultrathin films. We considered the features of alkaline ions, such as the ionic radius, polarizability, and electronegativity, to understand the changes in the structural, energetic, and electronic properties.

We found that the distortion of the octahedra is affected by the alkaline ionic radius, in which smaller adsorbed species lead to greater out-of-phase octahedra distortion and an increased band gap in systems with significant distortion. Among the systems studied, the \ce{Li} and \ce{Cs} systems showed the highest energetic stability, as indicated by their respective values of $E_{ad}$ and $E_{int}$. In the \ce{Li} system, sudden distortion of the \ce{PbI6}-octahedrs resulted in greater stabilization of nonbonding states \ce{I} 5p. On the other hand, in the \ce{Cs} system, the favorable acid-base interaction between \ce{Cs+} bases and \ce{I-} acids could explain the observed stability. 

We also investigated the effects of SOC and HSE06 on the electronic band gap of 2D perovskite slabs. Our results show that SOC reduces the electronic band gap, while HSE06 increases it. Interestingly, the use of the scissor operator to adjust the band gap values obtained from the PBE+D3 functional produced a good match, except for the case of \ce{Cs} on the monolayer. We also observed that the acid-base behavior of the $A$ site with the octahedra promotes a charge transfer from the alkaline ion to the iodide, which is more pronounced in systems with less electronegative alkaline ions.

\begin{acknowledgement}
The authors gratefully acknowledge the support from FAPESP (S{\~a}o Paulo Research Foundation, Brazil, Grant Numbers $2017/11631$-$2$ and $2018/21401$-$7$), Shell and the strategic importance of the support given by ANP (Brazil’s National Oil, Natural Gas and Biofuels Agency) through the R\&D levy regulation. I. C. R. thanks the National Council for Scientific and Technological Development (CNPq) for the Ph.D. fellowship, grant number $140015/2021$-$3$. A. F. B. B. also acknowledges financial support from FAPESP, grant number $2022/12778$-$5$ (postdoc fellowship). The authors are also thankful for the infrastructure provided to our computer cluster by the Department of Information Technology - Campus S{\~a}o Carlos.
\end{acknowledgement}

\begin{suppinfo}
The data employed for the figures, as well as additional analyses and technical details, are reported in the Supporting Information (SI). SI is available free of charge, in the online version at http://pubs.acs.org/
\end{suppinfo}

\bibliography{zboxref.bib}

\end{document}